\documentclass[sn-nature,Numbered]{sn-jnl}

\usepackage{graphicx}%
\usepackage{multirow}%
\usepackage{amsmath,amssymb,amsfonts}%
\usepackage{amsthm}%
\usepackage{mathrsfs}%
\usepackage[title]{appendix}%
\usepackage{xcolor}%
\usepackage{textcomp}%
\usepackage{manyfoot}%
\usepackage{booktabs}%
\usepackage{algorithm}%
\usepackage{algorithmicx}%
\usepackage{algpseudocode}%
\usepackage{listings}%
\usepackage{lineno}
\usepackage{aas_macros}
\usepackage{xcolor}
\usepackage{float}
\usepackage{orcidlink}

\newcounter{rtaskno}
\newcounter{rtaskno2}
\newcommand{\edf}[1]{\refstepcounter{rtaskno}\label{#1}}
\newcommand{\edt}[1]{\refstepcounter{rtaskno2}\label{#1}}

\theoremstyle{thmstyleone}%
\theoremstyle{thmstyletwo}%

\theoremstyle{thmstylethree}%

\begin{document}

\title[Article Title]{Fast-moving stars around an intermediate-mass black hole in  $\omega$\,Centauri}
\author*[1]{\fnm{Maximilian} \sur{Häberle \orcidlink{0000-0002-5844-4443}     }}\email{haeberle@mpia.de}
\author[1]{\fnm{Nadine} \sur{Neumayer \orcidlink{0000-0002-6922-2598}}}
\author[2]{\fnm{Anil} \sur{Seth \orcidlink{0000-0003-0248-5470}}}
\author[3]{\fnm{Andrea} \sur{Bellini \orcidlink{0000-0003-3858-637X}}}
\author[4,5]{\fnm{Mattia} \sur{Libralato \orcidlink{0000-0001-9673-7397}}}
\author[6]{\fnm{Holger} \sur{Baumgardt \orcidlink{0000-0002-1959-6946}}}
\author[2]{\fnm{Matthew} \sur{Whitaker \orcidlink{0009-0007-3015-9900}}}
\author[1]{\fnm{Antoine} \sur{Dumont \orcidlink{0000-0003-0234-3376}}}
\author[7,3]{\fnm{Mayte} \sur{Alfaro Cuello \orcidlink{0000-0002-1212-2844}}}
\author[3]{\fnm{Jay} \sur{Anderson \orcidlink{0000-0003-2861-3995}}}
\author[1,2]{\fnm{Callie} \sur{Clontz \orcidlink{0009-0005-8057-0031}}}
\author[8]{\fnm{Nikolay} \sur{Kacharov \orcidlink{0000-0002-6072-6669}}}
\author[9]{\fnm{Sebastian} \sur{Kamann \orcidlink{0000-0001-6604-0505}}}
\author[1,11]{\fnm{Anja} \sur{Feldmeier-Krause \orcidlink{0000-0002-0160-7221}}}
\author[10]{\fnm{Antonino} \sur{Milone \orcidlink{0000-0001-7506-930X}}}
\author[1]{\fnm{Maria Selina} \sur{Nitschai \orcidlink{0000-0002-2941-4480}}}
\author[9]{\fnm{Renuka} \sur{Pechetti \orcidlink{0000-0002-1670-0808}}}
\author[11]{\fnm{Glenn} \sur{van de Ven \orcidlink{0000-0003-4546-7731}}}

\affil[1]{\orgname{Max Planck Institute for Astronomy}, \orgaddress{\street{Königstuhl 17}, \postcode{69117 }\city{Heidelberg}, \country{Germany}}}

\affil[2]{\orgdiv{Department of Physics and Astronomy}, \orgname{University of Utah}, \orgaddress{\street{115 South 1400 East}, \city{Salt Lake City}, \postcode{84112}, \state{UT}, \country{USA}}}

\affil[3]{\orgname{Space Telescope Science Institute}, \orgaddress{\street{3700 San Martin Drive}, \city{Baltimore}, \postcode{21218}, \state{MD}, \country{USA}}}

\affil[4]{\orgdiv{AURA for the European Space Agency (ESA)}, \orgname{Space Telescope Science Institute}, \orgaddress{\street{3700 San Martin Drive}, \city{Baltimore}, \postcode{21218}, \state{MD}, \country{USA}}}

\affil[5]{\orgdiv{INAF}, \orgname{Osservatorio Astronomico di Padova}, \orgaddress{\street{Vicolo dell’Osservatorio 5}, \city{Padova}, \postcode{I-35122}, \country{Italy}}}

\affil[6]{\orgdiv{School of Mathematics and Physics}, \orgname{The University of Queensland}, \orgaddress{ \city{St. Lucia}, \postcode{4072}, \state{QLD}, \country{Australia}}}

\affil[7]{\orgdiv{Facultad de Ingenier\'{i}a y Arquitectura}, \orgname{Universidad Central de Chile}, \orgaddress{\street{Av. Francisco de Aguirre 0405}, \city{La Serena}, \state{Coquimbo}, \country{Chile}}}

\affil[8]{\orgname{Leibniz Institute for Astrophysics}, \orgaddress{\street{An der Sternwarte 16}, \city{Potsdam}, \postcode{14482}, \country{Germany}}}

\affil[9]{\orgdiv{Astrophysics Research Institute}, \orgname{Liverpool John Moores University}, \orgaddress{\street{146 Brownlow Hill}, \city{ Liverpool}, \postcode{L3 5RF}, \country{United Kingdom}}}

\affil[10]{\orgdiv{Dipartimento di Fisica e Astronomia “Galileo Galilei,”}, \orgname{Univ. di Padova}, \orgaddress{\street{Vicolo dell’Osservatorio 3}, \city{Padova}, \postcode{I-35122}, \country{Italy}}}

\affil[11]{\orgdiv{Department of Astrophysics}, \orgname{University of Vienna}, \orgaddress{\street{T\"urkenschanzstraße 17}, \city{Wien}, \postcode{1180}, \country{Austria}}}

\maketitle

\textbf{
Black holes have been found over a wide range of masses, from stellar remnants with masses of 5--150 solar masses (M$_\odot$), to those found at the centers of galaxies with $M>$10$^5$~M$_\odot$. 
However, only a few debated candidate black holes exist between 150 and 10$^5$ M$_\odot$.  Determining the population of these intermediate-mass black holes is an important step towards understanding supermassive black hole formation in the early universe  \cite{2020ARA&A..58..257G,2020ARA&A..58...27I}.
Several studies have claimed the detection of a central black hole in $\omega$\,Centauri, the Milky Way’s most massive globular cluster \cite{2008ApJ...676.1008N,2010ApJ...719L..60N,2017MNRAS.464.2174B}. However, these studies have been questioned due to the possible mass contribution of stellar mass black holes, their sensitivity to the cluster center, and the lack of fast-moving stars above the escape velocity \cite{2010ApJ...710.1032A,2010ApJ...710.1063V,2019MNRAS.482.4713Z,2019MNRAS.488.5340B}.  Here we report observations of seven fast-moving stars in the central 3 arcseconds (0.08 pc) of $\omega$\,Centauri. The velocities of the fast-moving stars are significantly higher than the expected central escape velocity of the star cluster, so their presence can only be explained by being bound to a massive black hole. From the velocities alone, we can infer a firm lower limit of the black hole mass of $\sim$8,200\,M$_\odot$, making this a compelling candidate for an intermediate-mass black hole in the local universe.}

\begin{figure}[h]%
\centering
\includegraphics[width=1.0\textwidth]{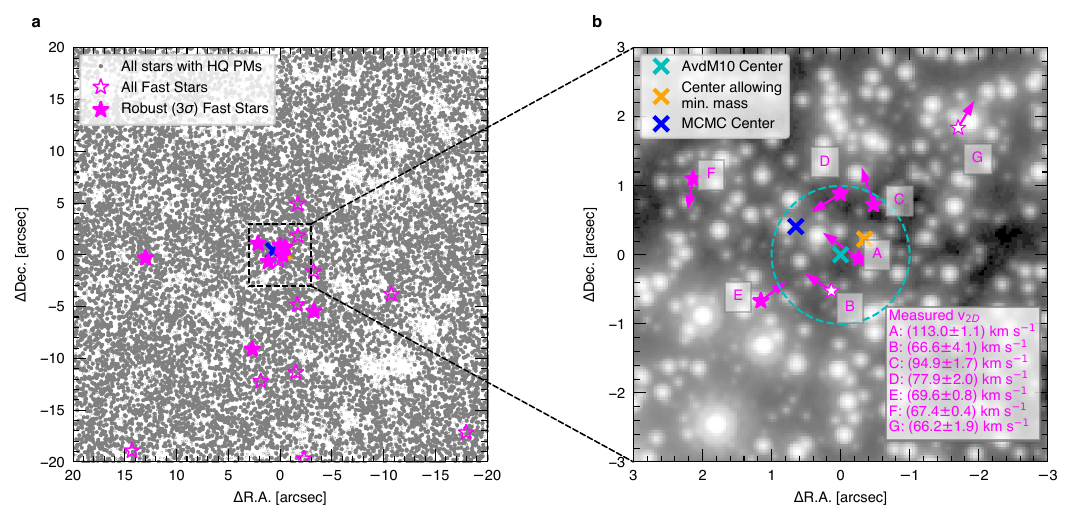}
\caption{\textbf{Location of fast-moving stars. }  \textbf{a},
This plot shows all stars we detected in our new proper motion catalog within a $40''\times40''$ region centered on the AvdM10 center \cite{2010ApJ...710.1032A}. Well-measured stars with velocities higher than the cluster escape velocity (62 km\,s$^{-1}$) and lying on the cluster main sequence in the color-magnitude diagram (Fig. \ref{fig:cmd}) are marked in pink. We use filled markers for stars that are at least 3$\sigma$ over the escape velocity. \textbf{b}, Stacked image of the innermost region of $\omega$\,Cen using all observations in the WFC3/UVIS F606W filter. The fast-moving stars and their proper motion vectors are shown in pink. The arrows indicate the stellar motion over 100 years. We also list the individual measured velocities in the lower right. The cyan cross indicates the photometric center of $\omega$\,Cen measured by \cite{2010ApJ...710.1032A} with the dashed circle indicating the 1" error reported for this center, the orange cross marks the center allowing for the lowest IMBH mass and the blue cross marks the most likely position of an IMBH given the MCMC analysis of the acceleration limits of the fast stars.
}\label{fig:position}
\end{figure}

$\omega$\,Centauri ($\omega$\,Cen) is a special case among the globular clusters of the Milky Way. Due to its high mass, complex stellar populations, and kinematics, $\omega$\,Cen is widely accepted to be the stripped nucleus of an accreted dwarf galaxy \cite{2000A&A...362..895H,2019NatAs...3..667I}. These factors combined with its proximity (D$=$5.43~kpc \cite{2021MNRAS.505.5957B}) have made it a prime target for searching for an IMBH.
As part of the \textit{oMEGACat} project\cite{2023ApJ...958....8N,2024arXiv240403722H}, we recently constructed an updated proper-motion catalog of the inner regions of $\omega$\,Cen, based on more than 500 \textit{Hubble Space Telescope} archival images taken over a timespan of 20 years. 
The unprecedented depth and precision of this catalog have allowed us to make the remarkable discovery of a significant overdensity of fast-moving stars in the center of the cluster (Fig.~\ref{fig:position} \& Extended Data Fig. \ref{edf:number_density}). In total, we find 7 stars with a total proper motion higher than 2.41\,mas\,yr$^{-1}$ within 3'' of the center determined in \cite{2010ApJ...710.1032A,2010AJ....140.1830G} (hereafter AvdM10 center). At a cluster distance of 5.43~kpc \cite{2021MNRAS.505.5957B}, this corresponds to projected 2D velocities higher than the escape velocity of the cluster if no IMBH is present  (v$_{\textrm{esc.}}$ = 62\,km\,s$^{-1}$ \cite{2018MNRAS.478.1520B}; see Methods).

\begin{figure}[h]%
\centering
\includegraphics[width=0.5\textwidth]{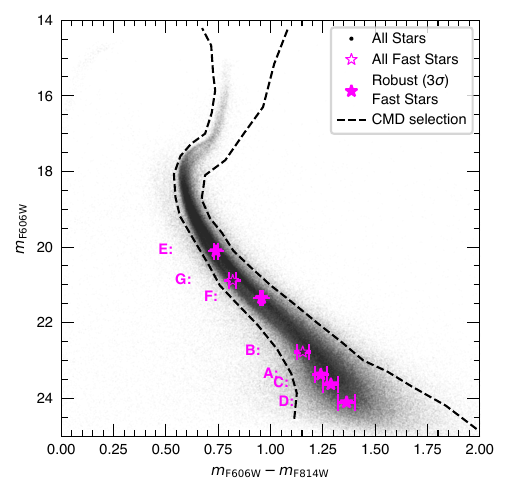}
\caption{\textbf{Hubble Space Telescope based color magnitude diagram (CMD) of $\omega$\,Cen.} The CMD locations of all fast-moving stars are marked with a pink symbol with their photometric 1$\sigma$ errors marked with error-bars. All of them lie on the main sequence, showing they are likely members of $\omega$\,Cen. The stars are labeled from A-G sorted by their distance from the AvdM10 center.}

\label{fig:cmd}
\end{figure}

\begin{figure}[h]%
\centering
\includegraphics[width=1.0\textwidth]{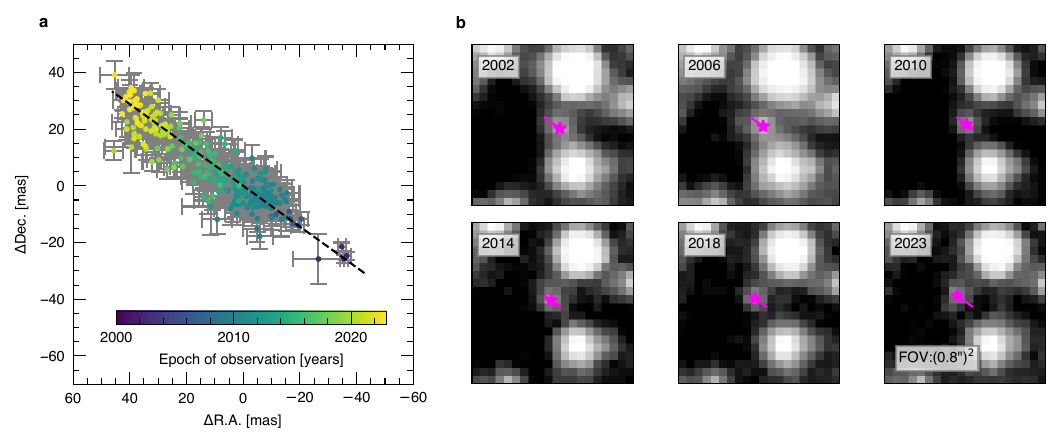}
\caption{\textbf{Motion of the fastest star. }  \textbf{a} Individual measured positions with 1$\sigma$ error-bars and \textbf{b} multi-epoch HST imaging  for Star A, the fastest ($v_{\textrm{proj.}}$ = 113.0$\pm$1.1 km\,s$^{-1}$) and centermost of the 7 newly discovered fast-moving stars in $\omega$\,Cen's center. The star is indicated with a pink marker on the images and its motion over 21 years with a line. This plot shows the excellent astrometric quality and the long temporal baseline we have in our unique dataset. Similar plots for all other stars are shown in Extended Data Fig. \ref{edf:multiepoch_imaging}.
}\label{fig2_fastest_star}
\end{figure}

We show in this paper that the presence of these stars strongly indicates a massive black hole, similar to the S-stars in the Galactic center \cite{2017ApJ...837...30G}.  A list of the fast-moving stars is shown in Extended Data Table \ref{edf:astrometry_table} and we label the fast-moving stars with letters from A--G, sorted by their proximity to the AvdM10 center. All these stars lie along the cluster main sequence in the color-magnitude diagram (Fig.~\ref{fig:cmd}). 
The fastest and centermost star (Star A in Fig.~\ref{fig:position}) has a 2D proper motion of 4.41$\pm$0.08\,mas\,yr$^{-1}$  (113.0$\pm$1.1\,km\,s$^{-1}$).
The motion of this star was measured over 286 epochs and a full 20.6~year time baseline (see Fig.~\ref{fig2_fastest_star}). 
We run extensive quality checks to ensure that the astrometry of the discovered fast stars is reliable. To ensure the cleanest possible dataset, we limit our analysis to stars whose velocity is at least 3$\sigma$ above the escape velocity. This leads to the exclusion of Stars B and G, however, this has negligible influence on the determined IMBH constraints.

Four of the fast stars, including the 3 fastest in the sample, are found within the centermost arcsecond ($r_{\textrm{projected}} <$ 0.03~pc or $ <0.09 $ ly). Surprisingly, these four innermost stars are all fainter than $m_{\textrm{F606W}}>22.7$, which is unlikely ($p=0.013$) to be a random occurrence given the overall distribution of stellar magnitudes in $\omega$\,Cen's center. In addition, all of them lie towards the blue side of the main-sequence. Both these properties could have interesting physical implications for the mechanism involved in capturing these stars or on their tidal interactions with the IMBH.

We expect a certain number of Milky Way stars in our field of view and because they have a large proper motion with respect to $\omega$\,Cen, they can mimic fast-moving cluster stars. Based on the number density of fast stars at larger radii (Extended Data Fig.~\ref{edf:number_density}), we estimate the rate of contaminants to be 0.0026\,arcsec$^{-2}$, which is consistent with expectations from the Besan\c{c}on Milky Way model \citep{2003A&A...409..523R}. This number density gives an expected average value of only 0.074 foreground stars in the inner 3-arcsecond radius. A detection of 5 such stars by a pure coincidence can therefore strongly be ruled out by simple Poisson statistics ($p = 1.7\times10^{-8}$) \cite{1986ApJ...303..336G}. Having 2 or more random contaminants within our 5-star sample can also be ruled out at the 3$\sigma$~level ($p=0.0026$). We also show in the Methods section that these stars cannot be explained by objects bound to stellar mass ($\lesssim$100~M$_\odot$) BHs, and that ejections from three-body interactions or an IMBH are not plausible.

Therefore, the presence of the seven central stars moving faster than the escape velocity of the cluster can only be explained if they are bound to a compact massive object near the center, raising the local escape velocity.  If no massive object was present, their velocities would cause them to leave the central region in less than 1,000 years, and then eventually escape the cluster. These fast stars are a predicted consequence of an IMBH, but are not expected from mass-segregated stellar-mass black holes \cite{2019MNRAS.488.5340B}.

We do not know several parameters of the system including the mass and exact location of this massive object, the relative line-of-sight distance between it and the stars, and the line-of-sight velocity of the stars. 
Despite this, we can calculate a lower limit on the mass of the dark object using the fast-moving stars' 2D velocities, assuming only that they are bound to it.  
The combined constraint from the 5 robustly measured fast stars' velocities is $\sim8,200$\,M$_\odot$, thus making an IMBH the only plausible solution.  The position of the IMBH requiring the lowest mass is only 0.3\,arcsec away from the AvdM10 \cite{2010ApJ...710.1032A} center, in excellent agreement with the $\pm$1" error on the AvdM10 center. Further details of this calculation can be found in the Methods.

While the linear motion and the velocity of the stars can be measured with great precision, the expected acceleration signal from an IMBH is considerably weaker and harder to detect. However, even a non-detection of acceleration could provide useful constraints on the mass and location of the IMBH. The accelerations of all stars are consistent with zero within 3$\sigma$, but two stars do have $>$2$\sigma$ acceleration measurements. We model both the velocity and acceleration measurements to further constrain the IMBH properties (see Methods); this calculation increases the lower limit on the black hole mass to 21,100~M$_\odot$ (99\% confidence) and gives a preferred position for the IMBH 0.77'' Northeast of the AvdM10 center.

In addition to these constraints that are purely based on the assumed escape velocity and our astrometric data of the 5 robustly measured fast-moving stars, we also compared the full velocity distribution observed in the inner 10'' of $\omega$\,Cen to already existing state-of-the-art N-Body models \cite{2017MNRAS.464.2174B} with various IMBH masses. Models with no IMBH,  a stellar mass black hole cluster, or with an IMBH with a mass greater than 50,000\,M$_\odot$ are all strongly ruled out, while models with an IMBH mass of 39,000\ and 47,000\,M$_\odot$ are most consistent with the fraction of fast stars and the observed velocity distribution. However, we caution that our comparisons show that low number statistics limit these comparisons, and mismatches with the overall velocity distribution suggest a need for improved modeling (see Methods for more details).

The detection of fast-moving stars in $\omega$\,Cen's center strengthens the evidence for an IMBH in this cluster. Due to $\omega$\,Cen's likely origin as the nucleus of the Gaia-Enceladus-Sausage dwarf galaxy \citep{2019A&A...630L...4M, 2021MNRAS.500.2514P} this black hole provides an important data point in the study of black hole demographics in low-mass galaxies, along with other black holes that have been detected in more massive globular clusters and stripped nuclei around M31 such as G1 (M$\sim20,000\,$M$_\odot$) \cite{2002ApJ...578L..41G,2005ApJ...634.1093G} or B023-G078 (M$\sim100,000\,$M$_\odot$) \cite{2022ApJ...924...48P}.
 In addition, this black hole provides the closest massive black hole and only the second after Sgr A* for which we can study the motion of multiple individual bound stellar companions. A comparison with the motion of the stars in the Galactic center is shown in the Methods and Extended Data Fig. \ref{edf:sstars}.

A more precise estimate of the black hole mass requires dynamical modeling of all newly available kinematic data using models that include the impact of both an IMBH and mass-segregated dark remnants. The exact properties of the orbits of the fast stars have to be determined by deep, pinpointed follow-up observational studies. Spectroscopic observations with integral-field-unit instruments such as VLT MUSE \cite{2010SPIE.7735E..08B} or JWST NIRSpec IFU \cite{2022A&A...661A..82B} could yield line-of-sight velocities for the fast-moving stars. Even more precise and deeper astrometric measurements with existing (VLTI GRAVITY+ \cite{2017A&A...602A..94G}, JWST NIRCam \cite{2023PASP..135b8001R}) or future (ELT MICADO \cite{2021Msngr.182...17D}, VLT MAVIS \cite{2021Msngr.185....7R}) instruments could enable the detection of additional tightly bound stars, and the measurements of accelerations, key for obtaining direct measurements of the black hole mass.
Our result also motivates revisiting the other likely accreted nuclear star clusters of the Milky Way \cite{2021MNRAS.500.2514P}, with M54 being the clearest case. For the search for IMBHs in other globular clusters, our results imply that it may be necessary to extend kinematic studies to the faintest stars, which is observationally challenging for clusters at larger distances and with high central densities.
\clearpage
\subsection*{Figures for main part}

\clearpage

\section*{Methods}

\subsection*{Discussion of previous IMBH detections in $\omega$\,Centauri}
The debate about an IMBH in $\omega$\,Cen dates back almost two decades but has remained controversial. Early dynamical modeling based on line-of-sight integrated-light velocity dispersion measurements suggested an IMBH of $M_{\textrm{IMBH}}=(4.0\,^{+0.75}_{-1.0})\times10^4$\,M$_\odot$ \cite{2008ApJ...676.1008N}.
These results were challenged with a precise redetermination of the center of the cluster \cite{2010ApJ...710.1032A} and dynamical modeling of proper motions \cite{2010ApJ...710.1063V} measured from multi-epoch  \textit{Hubble Space Telescope (HST)} imaging observations that placed an upper limit of $1.2\times10^4$\,M$_\odot$ on the IMBH. Using additional integrated light observations and a center based on the maximum LOS velocity dispersion, \cite{2010ApJ...719L..60N} obtained a best fit IMBH mass of $M_{\textrm{IMBH}}=(4.7\pm1.0)\times10^4$\,M$_\odot$. When assuming the AvdM10 center, the IMBH mass was slightly lower, $M_{\textrm{IMBH}}=(3.0\pm0.4)\times10^4$\,M$_\odot$.

Subsequent comparisons of both proper motions and line-of-sight velocities to N-Body simulations continued to show evidence for a $\sim4.0\times10^4$\,M$_\odot$ IMBH \citep{2012A&A...538A..19J,2017ApJ...842....6B}. However, these observations were shown to also be fully consistent with a dark cluster of stellar mass black holes in the central region of $\rm\omega\,Cen$ \cite{2019MNRAS.482.4713Z}. The lack of fast-moving stars in previous proper motion catalogs supported this scenario over an IMBH \cite{2019MNRAS.488.5340B}. Other works noted the influence of radial velocity anisotropy on dynamical mass estimates \cite{2017MNRAS.468.4429Z,2020MNRAS.499.4646A}.

Most recently, the discovery of a counter-rotating core using VLT MUSE line-of-sight velocity measurements of individual stars \cite{2024MNRAS.528.4941P} highlighted once again the kinematic complexity of the centermost region of $\omega$\,Cen. The center of this counter-rotation coincides with the AvdM10 center within $\sim$5", but is incompatible with the centers used in \cite{2008ApJ...676.1008N,2010ApJ...719L..60N}.

\subsection*{Previous Accretion Constraints in Context}
With the detection of a 10$^{4-5}$~M$_\odot$ IMBH, the upper limits on any accretion signal in the X-ray \cite{2013ApJ...773L..31H} and radio \cite{2018ApJ...862...16T} wavelengths make this the most weakly accreting black hole known. Deep $\sim$291 ksec Chandra observations place an upper limit of 0.5-7~keV luminosity of $\sim$10$^{30}$ ergs\,s$^{-1}$ \cite{2013ApJ...773L..31H}, roughly 12 orders of magnitude below the Eddington limit. The radio upper limit implies an even fainter source, with the 5~GHz upper limit of 1.3$\times$10$^{27}$ ergs\,s$^{-1}$ \cite{2018ApJ...862...16T} corresponding to an implied X-ray luminosity via the fundamental plane of $\sim$10$^{29}$ ergs\,s$^{-1}$ \cite{2012MNRAS.419..267P}.  Assuming standard bolometric corrections of $\sim$10 \cite{2020A&A...636A..73D}, this X-ray luminosity upper limit suggests an Eddington ratio log($L_{bol}/L_{edd}$)$< -12$, far fainter than that for Sgr~A* \cite{2022ApJ...930L..16E} or any other known black hole. This faint signal could be due to a combination of low surrounding gas density, a low accretion rate of that gas, and/or a low radiative efficiency \cite{2018ApJ...862...16T}.  Low-luminosity active galactic nuclei including Sgr~A* are brightest at IR and sub-mm wavelengths due most likely to synchrotron emission from compact jets \cite{2020A&A...638A...2G,2023A&A...670A..22F}.  Therefore, future observations with the James Webb Space Telescope or the Atacama Large Millimeter array would provide the highest sensitivity to any emission from $\omega$\,Cen's IMBH.  Any detection would reveal the location of the IMBH as well as provide valuable constraints on the black hole accretion in this extremely faint source.

\subsection*{Proper Motion Measurements and Sample Selection}
Our proper motion measurements are based on the reduction of archival \textit{Hubble Space Telescope} data of the central region of $\omega$\,Cen, taken over a time span of more than 20 years. We used the state-of-the-art photometry tool KS2 \cite{2017ApJ...842....6B} for the source detection and the astro-photometric measurements, and the established procedure described by \cite{2014ApJ...797..115B,2018ApJ...853...86B,2018ApJ...861...99L,2022ApJ...934..150L} to measure proper motions relative to the bulk motion of the cluster. The result of this extensive study is a proper motion catalog with high-precision measurements for 1.4 million stars out to $\omega$\,Cen's half-light radius with a typical temporal baseline of more than 20 years. 
Thanks to the large number of observations (in total we reduced over 500 images and some stars in the central region have up to 467 individual astrometric measurements) the catalog reaches unprecedented depth and precision. The highest precision is achieved in the well-covered center of the cluster, where our proper motions have a median error of only $\sim$6.6\,µas\,yr$^{-1}$ (0.17\,km\,s$^{-1}$) per component for bright stars.

The catalog is larger than any other kinematic catalog published for a globular cluster and significantly extends previous proper motion catalogs for $\omega$\,Cen \cite{2010ApJ...710.1032A,2017ApJ...842....6B,2023A&A...680A..35G}. A detailed comparison with other proper motion datasets is published along with the catalog \cite{2024arXiv240403722H}. In a following section and in Extended Data Fig. \ref{edf:photo_diagnostics} we compare the completeness of the different catalogs to show that it is plausible that the fast-moving stars have been missed in previous searches.

We use a high-quality subset of the proper motion catalog to search for real fast-moving stars and limit spurious astrometric measurements (e.g. two sources that are falsely identified as one) that can have apparent high proper motion measurements. Our criteria for this subset are based on the amount of available data for the measurements.  Specifically, we used only sources that had at least 20 astrometric measurements covering a temporal baseline of at least 20 years, and a fraction of rejected measurements (based on sigma clipping) of less than 15\%.  We also made cuts on the quality of the proper motion fit requiring both a proper motion error less than 0.194\,mas\,yr$^{-1}\approx$5\,km\,s$^{-1}$ and a reduced $\chi^2 < 10$ for the linear proper motion fit for both the R.A. and Dec. measurements. In addition to these quality selections, we also required the star to lie on the CMD sequence in an \textit{HST} based color-magnitude diagram (Fig.~\ref{fig:cmd}). These cuts all help to drastically reduce the number of contaminants.
These criteria are met uniformly out to a radius of $\sim$90 arcsec, at larger radii they lead to selection effects due to reduced observational coverage. 
A total of 157,320 out of 241,133 (65.2\%) entries of the proper motion catalog within $r<$90'' match the combined criteria.
Extended Data Table \ref{edf:astrometry_table}, b, shows the individual measured proper motion components for the 7 fast-moving stars. We note that for this analysis we have not applied the local a-posteriori proper motion corrections provided with the catalog \cite{2024arXiv240403722H}, as we are studying the central region which is well dithered and observed with various rotation angles. We verified that applying these corrections would neither change our fast star sample nor our conclusions.

\subsection*{Details on verification for fast-moving stars}
The criteria detailed above should lead to a clean data set with very few spurious proper motion measurements. To ensure that the measurements for the fast-moving stars are reliable, we inspected each of them carefully.

As a first step, we tested the quality of the raw astrometric measurements by studying several goodness-of-fit parameters and photometric quality indicators for the point-spread-function fits used to measure stellar positions (see Extended Data Fig. \ref{edf:photo_diagnostics}). We performed this analysis for the WFC3/UVIS F606W filter as it is the most used filter in the center of $\omega$\,Cen and each star has at least 195 measurements in this filter. To verify the goodness-of-fit, we used the mean of the so-called ``quality-of-fit" flag (QFIT) and the radial excess value, both of which take into account the residuals of the point-spread-function fit. In addition we looked at the mean of the ratio of source flux with respect to the flux of neighboring sources within its fit aperture. All five stars used for our analysis behave typically for well-measured stars of their magnitude and none of them show extreme values that would indicate problems with the photometry. It is noteworthy that the two stars excluded from our analysis based on their velocities being $<$3$\sigma$ above the escape velocity show some deviations: Star B has a relatively low mean QFIT value and high radial excess. Star G is the only one in the sample where the Flux Neighbour over Flux Star ratio is larger than one.

As a second step, we looked at the stars in several stacked \textit{HST} images taken at various epochs ranging from 2002 to 2023. The extensive multi-epoch imaging is demonstrated in Extended Data Fig \ref{edf:multiepoch_imaging}. The proper motion of a star at the escape velocity (62\,km\,s$^{-1}$) is 2.41\,mas\,yr$^{-1}$. Therefore, we expect to see a motion of at least 50\,mas over 21 years, which corresponds to 1.25 WFC3/UVIS pixels. Indeed this motion can be seen by eye for all seven stars in the multi-epoch images. Again the excluded stars B and G stick out in the sense that they are partially blended with neighboring stars, thus explaining their larger astrometric errors.

Finally, we tested the reliability of our proper motion measurements by limiting the raw position measurements to different subsets and redoing both the linear and quadratic fits to the motion of the stars. The first test run included only high S/N measurements. The second test run included only measurements taken with the WFC3/UVIS F606W filter. By using only one filter, we are immune to color-induced effects such as a partially resolved blend between two differently colored stars. The proper motions of all fast stars are consistent within the measurement uncertainties using both methods.

\subsection*{Comparison with other proper motion datasets}
Before our analysis, two other high-precision proper motion catalogs based on \textit{HST} data have been published \citep{2010ApJ...710.1032A,2017ApJ...842....6B} covering the center of $\omega$\,Cen. Both datasets were searched for central high proper motion stars, but none of the stars in our sample have been reported before. To understand why this is the case, we compare the completeness of the different catalogs. Extended Data Fig.~\ref{edf:photo_diagnostics}, \textbf{a} shows histograms of the magnitudes of stars with measured proper motions. In the inner 20'', our new catalog contains more than 3 times the number of stars of the literature catalogs and extends to significantly fainter magnitudes. The newly detected fast-moving stars all lie at faint magnitudes, where the completeness of the older catalogs is significantly lower than in the new proper motion catalog. This is due to larger amount of data and the updated source-finding algorithms in the new catalog, and explains the previous non-detection of the fast-moving stars.

\subsection*{Discussion of photometric errors}
Beside the astrometric reliability we also studied the quality of the photometric measurements used to locate the stars in the color-magnitude diagram. Even though the innermost stars are faint, their statistical photometric errors are small due to the large number of individual photometric measurements combined to a weighted mean value. The statistical errors range from 0.004 to 0.037 mag and are given in Extended Data Table~\ref{edf:astrometry_table}c. However, especially for faint stars, this statistical error is not able to capture systematic issues caused e.g. by the influence of brighter neighboring stars. Those can only be identified by verifying the quality of the PSF fit used to determine the individual photometric measurements. We report the mean quality-of-fit, radial excess, and ``neighbour flux / source flux" flags for both filters in Extended Data Table \ref{edf:astrometry_table}c and compare them with those of stars at similar ($\Delta m<0.5$) magnitudes in Extended Data Fig.~\ref{edf:photo_diagnostics}. All stars in the robustly measured sample show typical quality-of-fit for their respective magnitude. We note that stars B and G (which were excluded from the analysis) show comparatively poor QFIT. Star E and G show a possible flux contribution from a neighboring source (indicated by a high radial excess value and a high ``neighbor flux / source flux"). This can be confirmed by the stacked images shown in Extended Data Fig. \ref{edf:multiepoch_imaging}, where these stars show a close neighbor. Due to the relatively bright magnitude of star E and the low astrometric scatter we still consider its measurement valid.

\subsection*{Comparing the Density of Milky Way Contaminants to the Background}
To quantify our expected level of contamination from Milky Way foreground and background stars, we compared our results to those of a Besan\c{c}on model \citep{2003A&A...409..523R}.  Using the `m1612' model, we simulate a 1 square degree patch centered on $\omega$\,Cen retrieving Johnson colors and kinematics.  We then transform the model Johnson $V$ and $I$ magnitudes into F606W and F814W magnitudes using linear relations fitted to Padova models \cite{2012MNRAS.427..127B} between $V-I$ of 0 and 2.  We use the same color cuts and consider stars between F606W of 16 to 24. Then we count the number of stars with a total proper motion above 2.41\,mas\,yr$^{-1}$ (equivalent to our velocity cutoff at the escape velocity $v_{\textrm{esc.}}=$62\,km\,s$^{-1}$). These would appear as contaminants in our fast star sample. We find a density of 0.0039 stars\,arcsec$^{-2}$.  This is somewhat higher than the 0.0026$\pm$0.0003 stars\,arcsec$^{-2}$ found as the background level in our observations; this discrepancy is alleviated by considering only 65.2\% of all stars within our catalog meet the high-quality criteria used for our fast star selection (see above).  Correcting for this factor, we get an expected background of 0.0025 stars\,arcsec$^{-2}$, perfectly matching the observed background density (Extended Data Fig.~\ref{edf:number_density}).  This suggests that our background level is consistent with being predominantly Milky Way contaminants.  Relaxing our requirement that stars be more than 3$\sigma$ above the escape velocity results in a higher observed background level of 0.0042 stars\,arcsec$^{-2}$, no longer consistent with the Milky Way background. This suggests that our stricter definition of a fast-moving star reduces contamination from poorly measured stars in $\omega$\,Cen to a negligible level.

\subsection*{Discussion of other scenarios that could explain the fast-moving stars}
A complete contamination of our sample by Milky Way foreground/background stars that are non-members of $\omega$\,Cen can be ruled out statistically. We now explore and rule out alternative scenarios to the fast stars being bound to an IMBH. One alternative explanation for stars with a high velocity is to have them bound in a close orbit with a stellar-mass black hole. This scenario can be ruled out for BHs $<$100~M$_\odot$, as the periods required to reach the observed velocities are $<$10~years, well within the 20-year span over which we have observed linear motions. 

Another scenario could be that the stars are actually unbound from the cluster, and have been recently accelerated by three-body interactions, either with stellar-mass black hole binaries or an IMBH. Ejection by an IMBH in the center of the cluster can be ruled out by the very high rate of ejections necessary to sustain the observed number of fast stars within the center and the absence of observed fast-moving stars at larger radii.
To sustain a density of 0.18 fast stars\,arcsec$^{-2}$ in the inner 3 arcseconds (equivalent to our conservative sample of 5 stars) moving with at least 2.4~mas\,yr$^{-1}$, would require ejections with a rate of 0.004 stars\,yr$^{-1}$. This would lead to $\sim$117 additional fast-moving stars at larger radii (20"$ < r < $90"), in addition to $\sim$60 foreground stars expected from the (completeness corrected) Besan\c{c}on Milky Way model. In our dataset we find 61 fast-moving stars between 20"$<r<$90", consistent with the expected Milky Way background but not consistent with a significant number of additional ejected stars. In addition, a high hypothetical ejection rate of 0.004 stars\,yr$^{-1}$ would deplete all of $\omega$\,Cen's $\sim$10 million stars in just 2.5\,Gyr.  If no IMBH is present, accelerations of stars above the escape velocity are still possible by 3- or 4- body interactions between stellar or compact object binaries (see e.g. \cite{2023ApJ...946..104W}). However, these interactions would not be limited to the innermost few arcseconds of the cluster, due to the slowly varying stellar density in $\omega$\,Cen's core. In addition, the expected rate of these ejection events is of the order of less than one ejection per one million years, $\sim$1,000 times lower than needed to explain the observed number of fast stars in the center of $\omega$\,Cen \cite{2023ApJ...946..104W}\cite{2023ApJ...953...19C}.

\subsection*{Search for the fast-moving stars in recent line-of-sight velocity data}
Line-of-sight velocities of the fast-moving stars could help to exclude contaminants and provide additional constraints on the orbits of the stars and the mass and position of the IMBH. The deepest and most extensive spectroscopic catalog of stars in $\omega$\,Cen is Part I of our recently published \textit{oMEGACat} \cite{2023ApJ...958....8N}. This catalog was created using a large mosaic of observations with the VLT MUSE integral field spectrograph and contains both line-of-sight velocity measurements and metallicities for over 300,000 stars within the half-light radius of $\omega$\,Cen. While we could successfully cross-match 5 of the 7 fast-moving stars, their signal-to-noise ratio is typically too low for reliable velocity measurements, in particular for the 4 fastest, innermost stars ($S/N\sim2$).

We could, however, obtain a line-of-sight velocity value for star E ($v_{\textrm{LOS}}=261.7\pm2.7$\,km\,s$^{-1}$) and star F ($v_{\textrm{LOS}}=232.5\pm4.0$\,km\,s$^{-1}$). These velocities are very close to the systemic line-of-sight velocity of $\omega$\,Cen ($232.99\pm0.06$\,km\,s$^{-1}$ \cite{2023ApJ...958....8N}), confirming their membership in the cluster, as the Milky Way foreground is centered at $v_{\textrm{LOS}} \sim 0$ with a dispersion of 70\,km\,s$^{-1}$ \citep{2003A&A...409..523R}. 
However, as the relative line-of-sight velocity with respect to the cluster is low and we only have those two velocities for these outer stars, the line-of-sight velocities do not add stronger constraints on the IMBH. For this reason, we did not include them into the rest of our analysis.

\subsection*{Testing the robustness of the assumed escape velocity}
\textbf{Varying the parameters of the N-Body models:}
Because we use the escape velocity of $\omega$\,Cen (assuming no IMBH is present) as the threshold for determining whether a stars is considered ``fast" or not, it is important to verify the robustness of the escape velocity value.
We adopt an escape velocity v$_{\textrm{esc.}}$ = 62\,km\,s$^{-1}$ \cite{2018MNRAS.478.1520B}; we have verified this value based on fitting similar N-Body models to several state of the art datasets including MUSE LOS velocity dispersion measurements \cite{2018MNRAS.473.5591K} and \textit{HST} proper motion based dispersion measurements \cite{2015ApJ...803...29W} for the central kinematics and Gaia DR3 \cite{2023A&A...674A...1G} measurements at larger radii using an assumed distance of 5.43 kpc.
We varied both the assumed initial stellar mass function (using either the canonical Kroupa IMF \cite{2001MNRAS.322..231K} or the bottom-light IMF derived in \cite{2023MNRAS.521.3991B}) and the black hole retention fraction (assuming values of 10\%, 30\%, 50\%, or 100\%). Despite changes to the central M/L between models, the central escape velocity changes only minimally, with a range of values of from 61.1\,km\,s$^{-1}$ to 64.8\,km\,s$^{-1}$. Adopting any of these values leaves our sample of seven central stars above the escape velocity unchanged.

\noindent \textbf{An independent test using surface-brightness profiles}: As a second test, independent of the N-Body models, we calculated a surface brightness profile based escape velocity profile using various literature surface brightness profiles and dynamical models. We started by parameterizing the surface brightness profile using Multi-Gaussian Expansion (MGE) \cite{1994A&A...285..723E}  models. Then we converted the surface brightness to a mass density using several literature mass-to-light ratios and distances. From the mass density we can derive the gravitational potential ($\Phi(r)$). The escape velocity profile is then given by $v_{\textrm{esc.}}(r) = \sqrt{2(\Phi(r_{tidal},0) - \Phi(r,0))}$ (with the tidal radius $r_{tidal}= 48.6' \approx 74.6 {\textrm{pc}}$ from \cite{1996AJ....112.1487H,2010arXiv1012.3224H}).

These tests showed, that the central escape velocity does not depend strongly on the stellar mass distribution in the centermost region, instead it is dominated by the global M/L ratio and the assumed distance. The early dynamical models in the IMBH debate both assumed a distance of 4.8\,kpc \cite{2006A&A...445..513V} and a M/L of 2.6 (vdMA10, \cite{2010ApJ...710.1063V}) or 2.7 (N08, \cite{2008ApJ...676.1008N,2010ApJ...719L..60N}). With these values our tests give a central escape velocity of 55.4\,km\,s$^{-1}$ (vdMA10) and 56.9\,km\,s$^{-1}$ (N08). If we would also use the 4.8\,kpc distance to scale the proper-motions, this gives a cutoff of 2.43\,mas\,yr$^{-1}$ (vdMA10) and 2.48\,mas\,yr$^{-1}$ (N08), close to the adopted cut-off at 2.41\,mas\,yr$^{-1}$ and not changing the sample of seven detected fast stars. Thanks to the parallax measurements of the Gaia satellite and updated kinematic distance measurements, the distance to $\omega$\,Cen was robustly redetermined and larger values have been found (5.24$\pm$0.11\,kpc \cite{2021ApJ...908L...5S}; 5.43$\pm$0.05\,\cite{2021MNRAS.505.5957B}). A dynamical model using the same surface brightness profile as \cite{2008ApJ...676.1008N} but a larger distance of 5.14$^{+0.25}_{-0.24}$\, kpc was presented in \cite{2019MNRAS.482.4713Z}; this study found a M/L of 2.55$^{+0.35}_{-0.28}$. Using these values, the central escape velocity derived from the surface brightness profile is 61.1\,km\,s$^{-1}$  (equivalent to a proper motion of 2.51\,mas\,yr$^{-1}$); again not changing our fast stars sample. Finally, varying the distance of any model by 0.2 kpc while holding the M/L constant results in a $\sim$3 km/s variation in escape velocity. 
These results show that our fast star limit (v$_{\textrm{esc.}}$ = 62\,km\,s$^{-1}$ at a distance of 5.43\,kpc) is consistent with escape velocity values directly derived from surface-brightness profiles and several dynamically estimated M/L ratios. To visualize the escape velocity we calculated the escape velocity using the surface-brightness profile of \cite{2008ApJ...676.1008N},  a M/L ratio of 2.4, and a distance of 5.43\,kpc as found from the N-Body models with a cluster of stellar mass black holes. The resulting profile is shown in Extended Data Fig. \ref{edf:minmass}f. The predicted escape velocity is flat out to $\sim$50", a property shared by all of the calculated escape velocity profiles.  This makes the detection of the fast-stars only in the central few arcseconds more compelling.

\noindent \textbf{An empirical confirmation of the central escape velocity: }
We make one final, and relatively model-independent, empirical confirmation of the central escape velocity based on the distribution of 2D velocities in the innermost region of $\omega$\,Cen (see Extended Data Fig. \ref{edf:vesc_empirical}). As we have seen in the analysis above, the escape velocity only varies slightly within the inner $\sim50$" of the core of $\omega$\,Cen. In addition, the velocity dispersion profile is relatively flat in the innermost 10", with a value of $\sim$20 km/s \cite{2010ApJ...710.1032A,2015ApJ...803...29W,2024MNRAS.528.4941P}. Therefore, one would expect rather similar distributions of stellar velocities in both the very center ($0" < r < 3"$) and an outer ring at ($3" < r < 10"$). While we observe a clear excess of fast-moving stars in the inner 3 arcseconds, there is a sharp cutoff very close to the adopted escape velocity in the ($3" < r < 10"$) bin. Even though there is a total of 2090 stars there is only one star with a velocity significantly faster than the escape velocity (instead of ~17 stars expected from a 2D Maxwell-Boltzmann distribution with $\sigma_{\textrm{1D}}=20$\,km\,s$^{-1}$).  This suggests the stars with these velocities have escaped the central region. This one outer fast star has a 2D velocity of 75.8\,km\,s$^{-1}$ and is at a radius of $r=9.5"$. From the density of Milky Way contaminants with apparent velocities above the escape velocity we would expect $\sim$0.7 foreground stars in the  ($3" < r < 10"$) region, therefore this fast star is consistent with being a Milky Way foreground star. 

\subsection*{The Escape Velocity Provides a Minimum Black Hole Mass}
The escape velocity for an isolated black hole is given by
\begin{equation}
    v_{\textrm{esc., BH}}=\sqrt{\frac{2GM_{\textrm{BH}}}{r_{\textrm{3D}}}}.
\end{equation}
In $\omega$\,Cen we have to take into account the potential of the globular cluster as well. If we assume this to be constant over the very small region in which we found the fast-moving stars (an assumption that agrees with published surface brightness profiles, see Extended Data Fig. \ref{edf:minmass},\textbf{f}), we obtain:
\begin{equation}
    v_{\textrm{esc.,total}} = \sqrt{v_{\textrm{esc.,BH}}^2+v_{\textrm{esc.,cluster}}^2}
\end{equation}
If a star at the distance of $r_{\textrm{3D}}$ with a velocity $v_{\textrm{3D}}$ is bound to the black hole, we can calculate the following lower limit on the black hole mass\footnote{The original published version of this equation contained a typographical error. We have since issued a correction, see \href{https://doi.org/10.1038/s41586-024-08017-4}{DOI:10.1038/s41586-024-08017-4}. The arxiv version presented here [v3] has been corrected}:
\begin{equation}
    M_{\textrm{BH}}> \frac{(v_{\textrm{3D}}^2 - v_{\textrm{esc., cluster}}^2) \,r_{\textrm{3D}}}{2G}\geq \frac{(v_{\textrm{2D}}^2 - v_{\textrm{esc., cluster}}^2) \,r_{\textrm{2D}}}{2G}
\end{equation}
A lower limit can also be calculated if the line-of-sight velocity and distance are not known, as $v_{\textrm{3D}}\geq v_{\textrm{2D}}$ and $r_{\textrm{3D}}\geq r_{\textrm{2D}}$. Since we do not know the exact 2D position of the black hole relative to the fast-moving stars, we calculated this lower limit for all stars and a grid of assumed 2D locations around the AvdM10 center \cite{2010ApJ...710.1032A}. Each individual star alone would allow for a very low mass, as the location of the black hole could coincide with the star (Extended Data Fig. \ref{edf:minmass}, a-d). However, combining these limits for all stars gives a higher minimum black hole mass  (Extended Data Fig. \ref{edf:minmass}, e). If we assume that all 5 robustly detected stars are bound to the black hole, the lower limit is $\sim8,200$\,M$_\odot$ and the minimum mass location is only 0.3" away from the AvdM10 \cite{2010ApJ...710.1032A} center at the location R.A.: 201.6966908$^\circ$\footnote{This value has been corrected for an error that occurred during the conversion from relative to absolute coordinates. See also \href{https://doi.org/10.1038/s41586-024-08017-4}{DOI:10.1038/s41586-024-08017-4}} Dec.:-47.4795066$^\circ$. If we assume that the 2 most constraining stars are just random foreground contaminants, which is ruled out at the 3$\sigma$ level ($p=0.0026$), this limit drops to $\sim4\,100$\,M$_\odot$, still well within the IMBH range.

\subsection*{Acceleration Measurements}
The astrometric analysis in the catalog \cite{2024arXiv240403722H} considered only linear motions of the stars. If there is a massive black hole present near the center, we might also be able to measure accelerated motion of the closest stars, allowing for a direct mass measurement of the black hole. With an IMBH mass of 40,000\,M$_\odot$ and at a radius of 0.026\,pc (1'' on the sky), the acceleration of a star would be 0.25\,km\,s$^{-1}$\,yr$^{-1}$ (or 0.01\,mas\,yr$^{-2}$). This is at the limit of the precision of our current dataset: With a 20 year baseline, we only expect a deviation of 0.05 pixel from a linear motion. For bright stars the astrometric uncertainty can be as low as 0.01\,pixel, however for the faint fast-moving stars we have detected, the errors are significantly larger. 

To constrain these possible accelerations, we repeated the fit of the motion of each star allowing the addition of a quadratic component. The results for this fit are shown in Extended Data Table \ref{edf:astrometry_table}.
All of the stars' accelerations are consistent with zero within 3$\sigma$, but two stars have $>$2$\sigma$ acceleration measurements. The errors on our acceleration measurements lie between 0.004 and 0.03\,mas\,yr$^{-2}$ and are, therefore, of a magnitude similar to the expected acceleration signal. The strongest acceleration is shown by Star B, which has been excluded from the robust subset of fast-moving stars because its proper motion is not 3$\sigma$ above the escape velocity. Due to the proximity of a bright neighbor star, we do not deem this acceleration measurement to be reliable.

As the line-of-sight distances to the fast stars are unknown, it is not possible to place direct constraints on the IMBH mass using the upper limits on accelerations. If no acceleration is detected, as is the case for the centermost star A, this could mean that either the black hole is not very massive or that the line-of-sight distance of star~A to the black hole is large. Combining the measurements for the ensemble of fast-moving stars, and making some assumptions on their spatial distribution still allows us to use the acceleration limits to place additional constraints on the black hole mass and its location. This is described in the next section.

\subsection*{Details about MCMC fitting of the acceleration data}
Assuming the fast-moving stars are bound to the IMBH, we can model the stars as being on Keplerian orbits around the IMBH.  We used Bayesian analysis to sample the posterior distribution for the unknown mass and position of the black hole. In this analysis, there were 8  free parameters: black hole mass, its on-sky x- and y-position, and 5 line-of-sight distances between the black hole and each fast-moving star. This analysis makes use of the available astrometric observations but stops short of modeling individual stellar orbits which would introduce additional free parameters.

We use a likelihood function with these 8 free parameters and give the likelihood based on the observed on-sky x- and y- acceleration, proper motion, and position of the 5 robustly measured fast-moving stars. For each star, we calculated a first likelihood term based on the modeled acceleration $a_{\textrm{modeled}}$ using a Gaussian distribution with mean $a_{\textrm{observed}}$ 
and width equal to the acceleration uncertainty. The second term in the likelihood accounts for the escape velocity constraints and is kept constant if the observed 2D velocity of the star is below the modeled escape velocity. For stars with 2D velocities above the modeled escape velocity, the likelihood is a Gaussian distribution with mean $v_{\textrm{2D}} - v_{\textrm{esc. total}}$ and width equal to the uncertainty in observed proper motion.

We make these prior assumptions about the model: 1) The black hole mass is between $1\,\mathrm{M_\odot}$ and $100,000\,\mathrm{M_\odot}$, since a black hole mass beyond this upper limit is ruled out by our N-Body models. 2) The black hole is located within the distribution of the fast-moving stars. We use a Gaussian prior in the black hole x- and y- positions with a mean equal to the mean position of the fast-moving stars and width equal to their 1-dimensional positional standard deviation, $\sigma_{\textrm{stars}} = 0.0221\,\mathrm{pc}$; we also use a cutoff at $\pm 0.16\,\mathrm{pc}$. 3) The stellar positions are isotropically distributed around the black hole. We model the line-of-sight positions of the stars relative to the black hole using a Gaussian distribution with mean 0 and width $\sigma_{\textrm{stars}}$.

The posterior was sampled using a Markov chain Monte Carlo (MCMC) ensemble sampler implemented using the package \textit{emcee} \cite{2013PASP..125..306F} using recommended burn-in and autocorrelation corrections. 
We show the posterior distribution for the black hole mass in Extended Data Fig. \ref{edf:accelerations}, a.   The 99\% confidence lower limit (21,100~M$_\odot$) is significantly higher than that derived from escape velocity constraints alone, while the upper limit on the mass is not well constrained. We also find a position for the black hole east of the AvdM10 center, with $\Delta x = -0.017\,^{+0.017}_{-0.031}\,\mathrm{pc}$ and $\Delta y = 0.011\,^{+0.011}_{-0.025}\,\mathrm{pc}$ (Extended Data Fig. \ref{edf:accelerations}, b. The coordinates of the MCMC based center estimate are R.A: 201.6970988$^\circ$\footnote{This value has been corrected for an error that occurred during the conversion from relative to absolute coordinates. See also \href{https://doi.org/10.1038/s41586-024-08017-4}{DOI:10.1038/s41586-024-08017-4}} Dec.: -47.4794533$^\circ$. We note that the black hole location estimate is dominated by the marginal 2$\sigma$ acceleration signal of Star D which is the faintest star in the sample; follow-up studies are required to obtain more precise acceleration measurements.

\subsection*{N-Body Models}
In addition to the analysis of stars with velocities above the escape velocity, we also used a set of existing N-body
models with and without central IMBHs to get additional constraints on the IMBH mass.  We compared the simulations to the full velocity dispersion and surface density profile of $\omega$\,Cen to determine the best-fitting model and the mass of a central IMBH. The set of models and the details of the fitting procedure are described in detail in \cite{2017MNRAS.464.2174B,2019MNRAS.488.5340B}. We note that these models have been presented already in the literature, but the fits to these models have been updated to incorporate the most recent Gaia DR3 data.

In short, the models started from King profiles \cite{1962AJ.....67..471K} with central concentrations between $c=0.2$ and $c=2.5$ and initial half-mass radii between $r_h=2$ pc and $r_h=35$ pc. In the models with an IMBH, we varied the mass of the IMBH so that it contains either 0.5\%, 1\%, 2\% or 5\% of the cluster mass at $T=12$ Gyr when the simulations were stopped. The models with an IMBH assumed a retention fraction of stellar-mass black holes of 10\% while in the models without an IMBH we varied the assumed retention fraction of stellar-mass black holes between 10\% to 100\%. At the end of the simulations, we calculated surface density and velocity dispersion profiles for each $N$-body model and then determined the best-fitting model by interpolation in our grid of models and using $\chi^2$ minimization against the observed velocity and surface density profile of $\omega$ Cen. 

The velocity distributions from observations and the models are shown in Extended Data Fig. \ref{edf:nbody}. We compare the distribution of measured 2D stellar velocities in the inner 10'' of $\omega$\,Cen with the various models using a Kolmogorov–Smirnov test. In addition, we compare the fraction of fast-moving stars in the innermost 3 arcseconds (see Extended Data Table \ref{edf:nbody_table}).
Models without an IMBH and with a 20,000\,M$_\odot$ IMBH are both strongly excluded both by the overall velocity distribution and the complete lack of fast-moving stars.
The overall velocity distribution is in best agreement with the 47,000\,M$_\odot$ distribution, while the fraction of fast stars is best matched by the 39,000\,M$_\odot$ simulation. This tension might be alleviated in future models, that contain both an IMBH and a cluster of stellar mass black holes. We caution that these simulations have smaller numbers of stars than observed, and that there can be significant variations in the distribution of central stars due to strong encounters with remnants and binaries. Nonetheless, the simulations suggest black holes with masses of $M\lesssim$50,000\,M$_\odot$ are consistent with the observed distribution of central velocities and fast-moving stars, while the no-IMBH case and significantly more massive black holes are disfavored due to an overprediction of fast stars. Updated N-Body models fit to the oMEGACat kinematic data and dynamical modeling of these same datasets with Jeans models are currently underway.

\subsection*{Comparison with S-Stars in the Galactic center}
The black hole indicated by our fast star detection is only the second after Sgr A*, for which we can study the motion of multiple individual bound stellar companions. Therefore, the extensively measured stars around Sgr A* provide a unique comparison point to our fast-moving star sample. We compare the motions of the stars in the S-Star catalog from Gillessen 2017 \cite{2017ApJ...837...30G} with our $\omega$\,Cen fast star sample in Extended Data Fig. \ref{edf:sstars}. When taking into account the different distances and the approximate black hole mass ratio of 100, the motions indeed show similar amplitudes. However, the density of tracers in $\omega$\,Cen is significantly lower, despite the greater depth of the observations.

\clearpage
\backmatter

\bmhead{Acknowledgments}

Based on observations with the NASA/ESA Hubble Space Telescope, obtained at the Space Telescope Science Institute, which is operated by AURA, Inc., under NASA contract NAS 5-26555. SK acknowledges funding from UKRI in the form of a Future Leaders Fellowship (grant no. MR/T022868/1). AS, MAC, MW, and AB acknowledge support from HST grant GO-16777. AB acknowledges support from STScI grants GO-15857 and AR-17033. AFK acknowledges funding from the Austrian Science Fund (FWF) [grant DOI 10.55776/ESP542]

\section*{Declarations}
\subsection*{Data availability}
The data used in this paper are based on archival observations taken with the \textit{Hubble Space Telescope} which are freely available in the Mikulski Archive for Space Telescopes. All used observations have been grouped under a DOI \cite{mydoi}. In addition, the full proper-motion catalog is made public along with the respective publication \cite{2024arXiv240403722H}.

\subsection*{Code availability}
We used the following python packages to perform the analysis: 
\texttt{matplotlib \cite{2007CSE.....9...90H}, scipy \cite{2020NatMe..17..261V}, numpy \cite{2020Natur.585..357H}, astropy \cite{2022ApJ...935..167A}, emcee \cite{2013PASP..125..306F}}. The N-Body simulations were run with the publicly available NBODY6 code \cite{2012MNRAS.424..545N}. We can share the code used in the data analysis upon request.
\subsection*{Author's contributions}
All authors helped with interpretation of the data and provided comments on the manuscript. MH has led the analysis of the data and is the main author of the text. AS and NN designed the overall project with significant contributions from AB, GvdV, and SK.  AS, NN, HB, MW, and AD contributed text. AS determined the expected density of Milky Way contaminants. AB, ML and JA provided their expertise on astrometric measurements with the \textit{HST}. HB provided and fitted the N-Body models for the analysis. MW ran the Bayesian Analysis used to constrain the IMBH mass and position. AD performed the surface-brightness-profile based calculations. SK and MSN helped to find available LOS data for the stars.
\subsection*{Competing interests}
The authors declare no competing interests.

\bibliography{sn-bibliography}

\clearpage
\section*{Extended Data}
\edt{edf:astrometry_table}
\small\noindent\textbf{Extended Data Table~\ref{edf:astrometry_table}: Detailed astrometric and photometric information of the fast-moving stars.}
The table in \textbf{a} shows the position, the number of astrometric measurmements $N_{\textrm{used}}$, and the projected distance from the AvdM10 center for each of the fast-moving stars. \textbf{b} lists the individual proper motion components, the total proper motion, the inferred 2D velocity and the measurements of the acceleration for the seven fast-moving stars. All shown errors correspond to the 1$\sigma$ errors, which were estimated by scaling the formal errors on the parameters by $\sqrt{\chi^2_{red.}}$ of the respective fit. The strongest accelerations are shown by Star B, however, this star has been discarded from the set of robustly measured stars due to large astrometric errors. All robust stars show an acceleration consistent with zero to within 2 $\sigma$. Finally, Table \textbf{c} lists several photometric properties of the fast-moving stars in two filters, including the measured brightness and several photometric diagnostics (the quality of fit (QFIT) parameter, the radial excess (RADX) parameter and the flux ratio between the flux of each source and the flux of neighboring sources). Together with the photometric diagnostic values we show their percentile with respect to stars with a similar magnitude ($p$).
\vspace{2cm}

\begin{figure}[h!]%
\centering
\includegraphics[trim={0 5cm 0 9cm},width=1.0\textwidth]{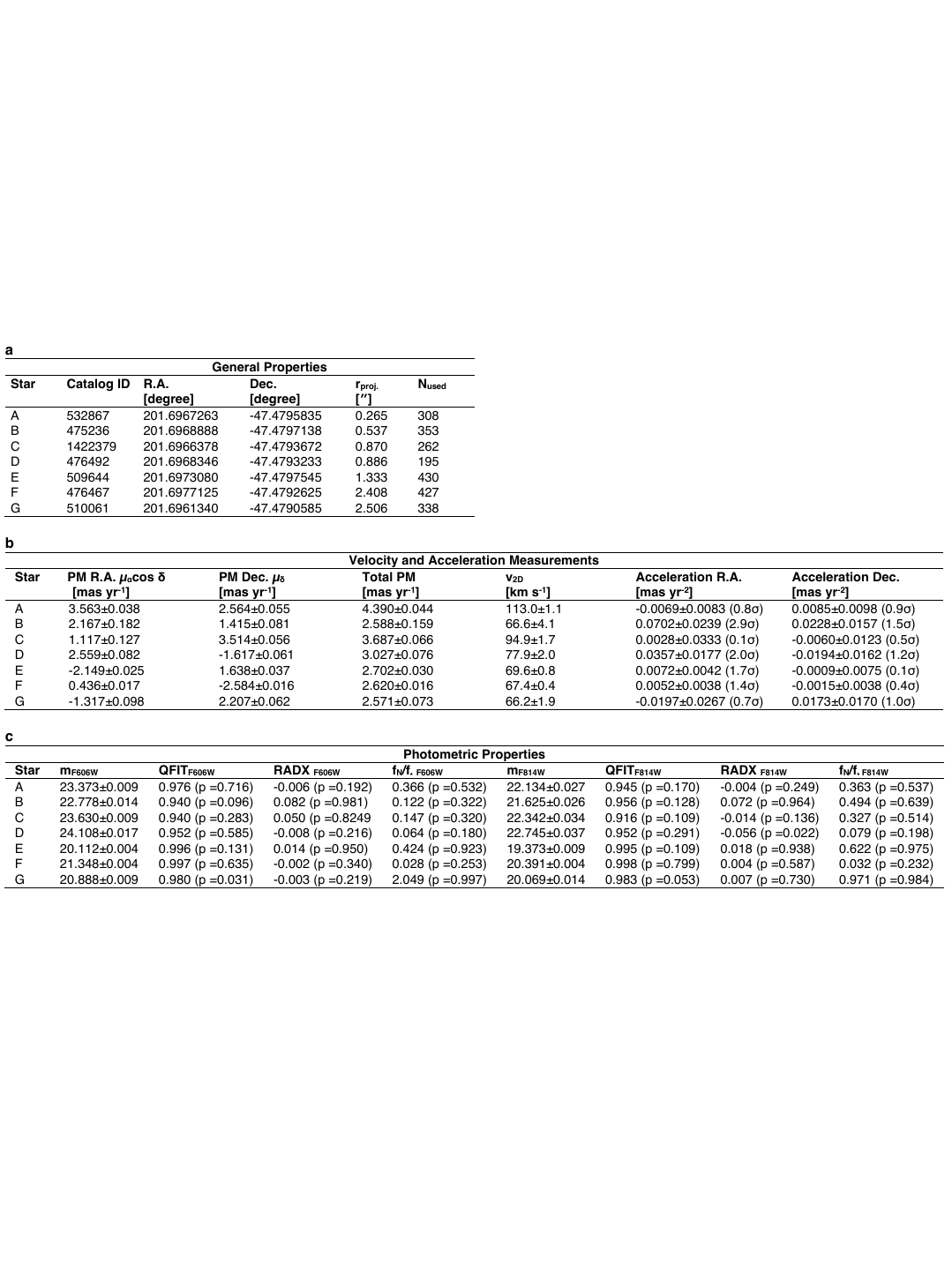}
\end{figure}
\clearpage
\edf{edf:number_density}
\small\noindent\textbf{Extended Data Fig. \ref{edf:number_density}: Number density of fast-moving stars.}
The black markers with errorbars (1$\sigma$; based on Poisson statistics) show the measured number density of robustly detected fast-moving stars determined in radial bins with respect to $\omega$\,Cen's center. A constant density of Galactic fore-/background stars is expected, but we see a strong and statistically significant rise of the density towards the AvdM10 \cite{2010ApJ...710.1032A} center.  Based on the number density of fast-moving stars at large radii (which is consistent with predictions from a Besan\c{c}on Milky Way model, dashed line) only $\sim$0.073 fast-moving stars are expected within the central 3” compared to the 5 observed stars.
\begin{figure}[h]%
\centering
\includegraphics[width=0.5\textwidth]{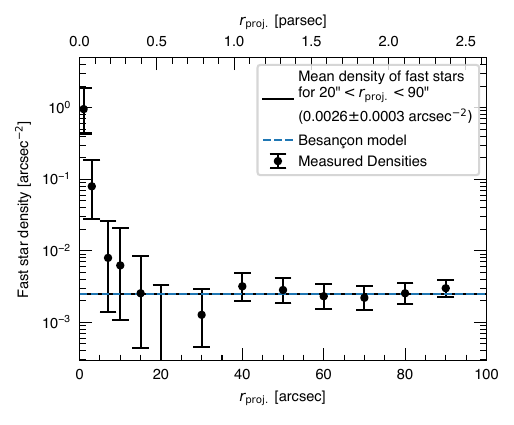}
\end{figure}
\clearpage

\edf{edf:multiepoch_imaging}
\small\noindent\textbf{Extended Data Fig. \ref{edf:multiepoch_imaging}: Astrometry and multi epoch imaging for all fast-moving stars.}
Each row (\textbf{a-g}) shows the astrometry and imaging for one of the 7 fast-moving stars. The left column shows the raw astrometric measurements used to determine the proper motions, color-coded by the epoch of their observation. The center column shows the linear and quadratic fits to both the R.A. and Dec. position change of the stars. Errorbars correspond to the 1$\sigma$ error on the individual position measurements. The right column shows stacked images from 2002 (ACS/WFC F625W) and 2023 (WFC3/UVIS F606W), the positions of the fast stars are marked with a pink open circle.
\begin{figure}[h!]%
\centering
\includegraphics[width=0.8\textwidth]{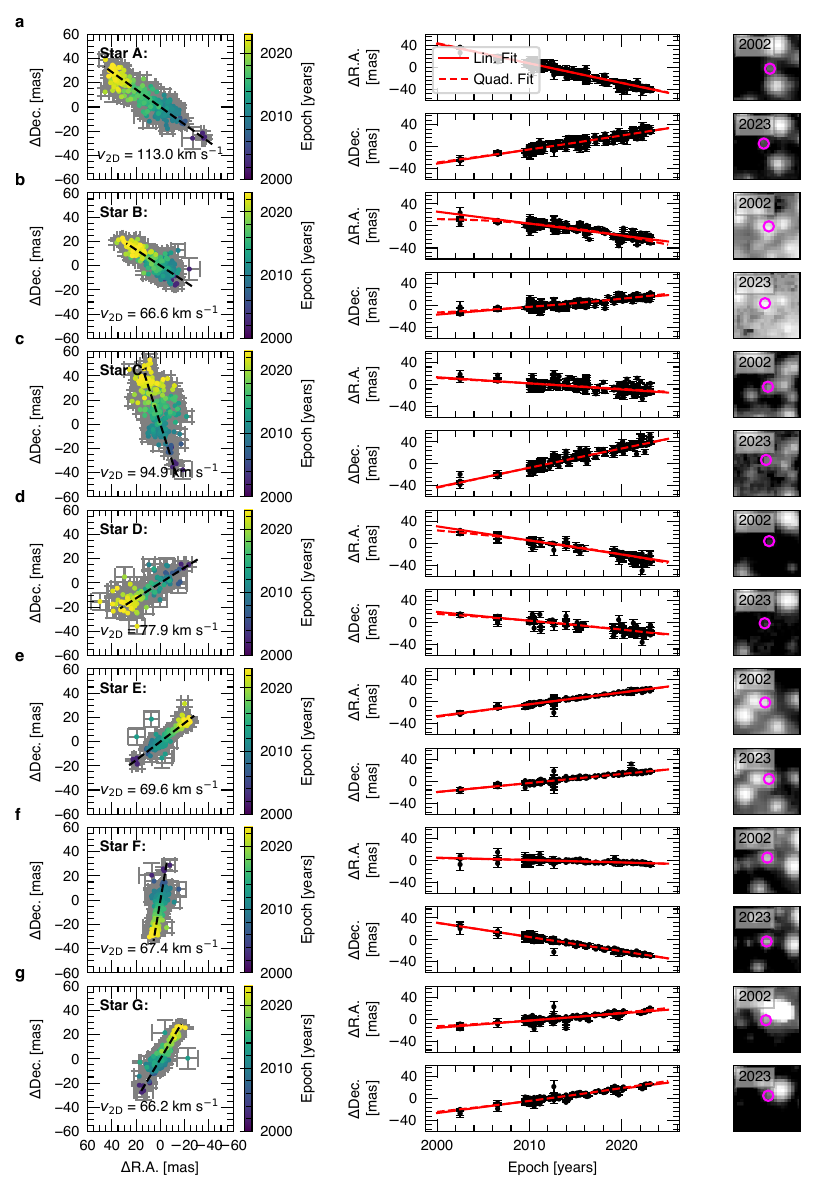}
\end{figure}

\clearpage
\edf{edf:photo_diagnostics}
\small\noindent\textbf{Extended Data Fig. \ref{edf:photo_diagnostics}: Completeness of the catalog and photometric diagnostics.}
Panel \textbf{a} compares the completeness of the various available proper motion datasets for the core of $\omega$\,Cen using histograms of the magnitude distribution. The new oMEGACat \cite{2024arXiv240403722H} has significantly higher completeness and reaches fainter magnitudes than the literature catalogs even if we apply the strict quality criteria used in this work. This explains why previous catalogs have not found the faint fast-moving stars (marked with vertical lines) we detect here.

The grey dots in \textbf{b},  \textbf{c}, and  \textbf{d} show the mean of photometric diagnostics for the raw PSF photometry measurements for the fast-moving stars compared with the bulk of stars in the catalog. The first panel shows the QFIT parameter, given by the linear correlation function between the PSF and the pixel values in the image. The second panel shows the radial excess parameter, a parameter that compares the residual flux inside versus outside of the fit aperture. The third panel shows the flux ratio between the flux of a star itself and its neighboring sources. All 5 robustly measured stars show typical behavior for their magnitude, while the excluded fast-moving stars (B, G) are influenced by bright neighboring sources.
\begin{figure}[h!]%
\centering
\includegraphics[width=0.9\textwidth]{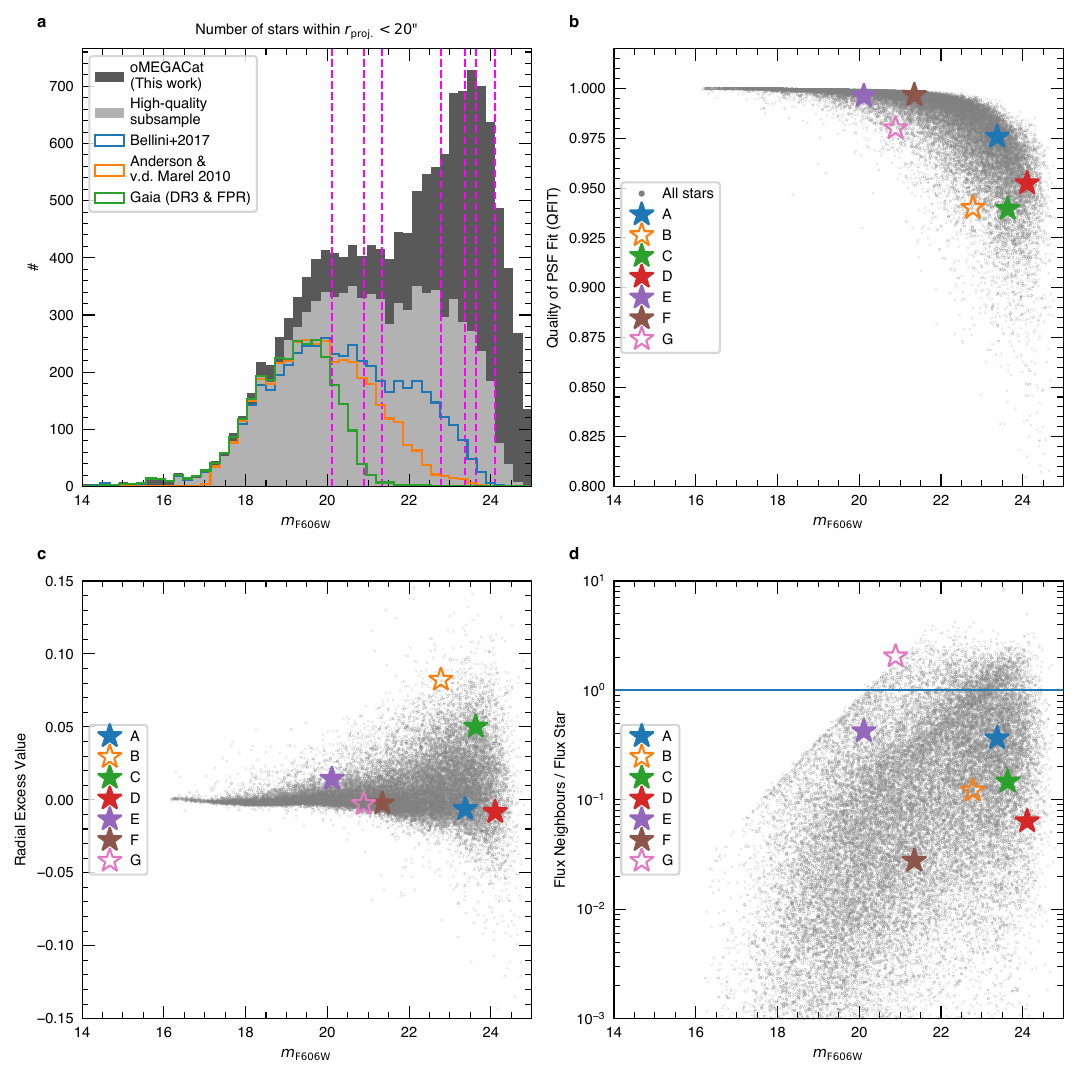}
\end{figure}

\clearpage
\edf{edf:vesc_empirical}
\small\noindent\textbf{Extended Data Fig. \ref{edf:vesc_empirical}: Empirical verification of the escape velocity.}
Panels \textbf{a} and \textbf{b} show histograms of the observed 2D velocity distribution in the very center ($0" < r<3$"; \textbf{c}) and in an outer ring ($3" < r < 10"$; \textbf{d}). While the lower velocities are well described by a 2D Maxwell-Boltzmann distribution with $\sigma_{\textrm{1D}}=20$~km\,s$^{-1}$ (marked with a solid black line, the dashed black lines refer to alternative distributions with $\sigma_{\textrm{1D}}=17$~km\,s$^{-1}$ and $\sigma_{\textrm{1D}}=23$~km\,s$^{-1}$), there are clearly notable differences at higher velocities. Those become especially visible in the cumulative normalized histogram shown in Panel \textbf{d} and the zoom-in in Panel \textbf{e}: While the distribution between ($0" < r<3$", blue line) shows an excess of fast-moving stars, the distribution at larger radii ($3" < r < 10"$, orange line) shows a clear deficit of stars at velocities larger than the escape velocity, making the used escape velocity threshold very plausible. Even though the sample is 10 times larger, there is only a single star with a velocity significantly larger than $v_{\textrm{esc.}}$. This star has a 2D velocity of 75.8\,km\,s$^{-1}$ and is at a radius of $r=9.5"$. It is consistent with being a Milky Way foreground star.
\begin{figure}[h!]%
\includegraphics[width=1.0\textwidth]{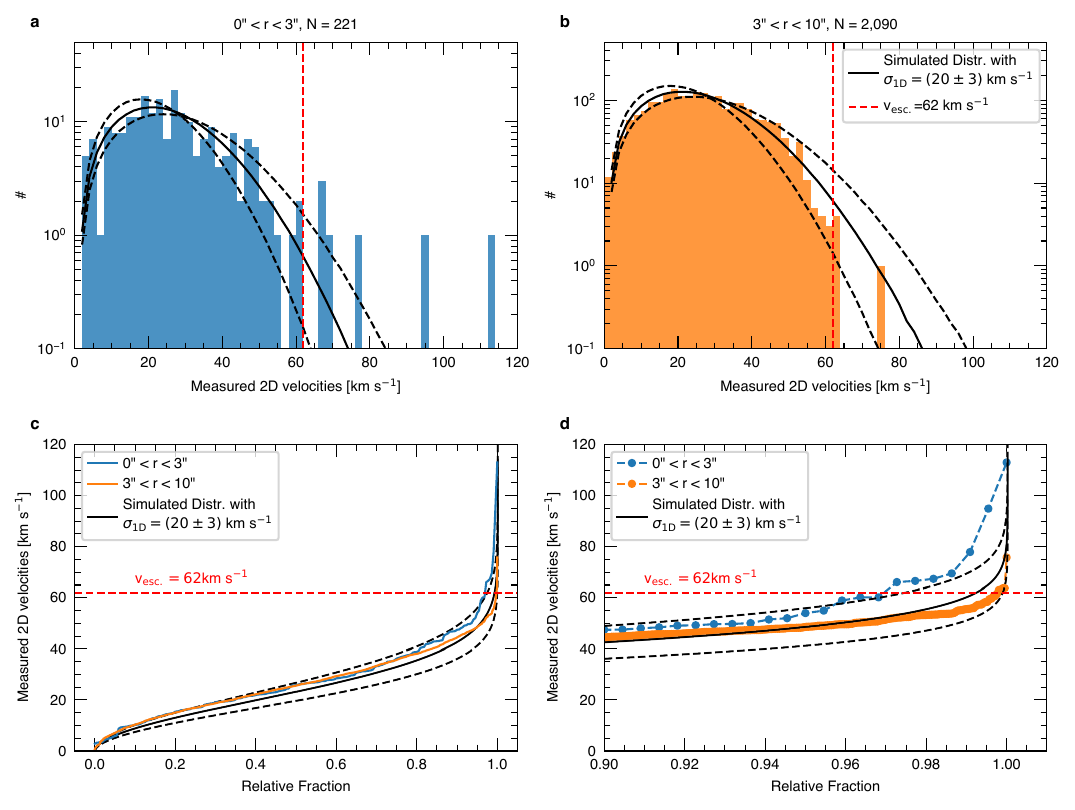}
\end{figure}

\clearpage
\edf{edf:minmass}
\small\noindent\textbf{Extended Data Fig. \ref{edf:minmass}: Determination of a lower limit on the IMBH mass using the escape velocity. }
The presence of stars with velocities above the escape velocity of the cluster indicates that they are bound to a massive object. Since we neither know the mass nor the exact position of the object, we can only infer a lower mass limit for each possible 2D location. The 4 left plots (\textbf{a-d}) show the contours of the minimum black hole mass indicated by the 4 centermost robustly measured fast-moving stars. By combining the minimum black hole mass constraints from each of the fast-moving stars we can find the position that allows for the lowest IMBH mass (\textbf{e}). This analysis indicates a firm lower limit of around 8,200 M$_\odot$. Our minimum mass location only differs by $\sim$0.3 arcsec from the AvdM10 \cite{2010ApJ...710.1032A} center. This result does not significantly change if we assume some of the fast-moving stars are contaminants and remove them from the analysis.

Panel \textbf{f} shows the results for a surface brightness profile based escape velocity profile in blue, using the surface brightness profile from \cite{2008ApJ...676.1008N} and the dynamical distance (5.43\,kpc) and M/L ratio (2.4) derived from N-Body models, either without any IMBH, with a 8,200\,M$_\odot$ IMBH, or a 40,000\,M$_\odot$ IMBH. The radii are measured with respect to the minimum mass center shown in \textbf{e}. The shaded regions indicate the uncertainty introduced by an assumed error of $\pm$0.2\,kpc on the distance. The surface-brightness profile based escape velocity is compatible with the adopted value of $v_{\rm}=$62\,km\,s$^{-1}$. The profile without an IMBH is also nearly flat in the inner $\sim$50", justifying the assumption of a flat profile in the innermost region.

\begin{figure}[h!]%
\centering
\includegraphics[width=0.85\textwidth]{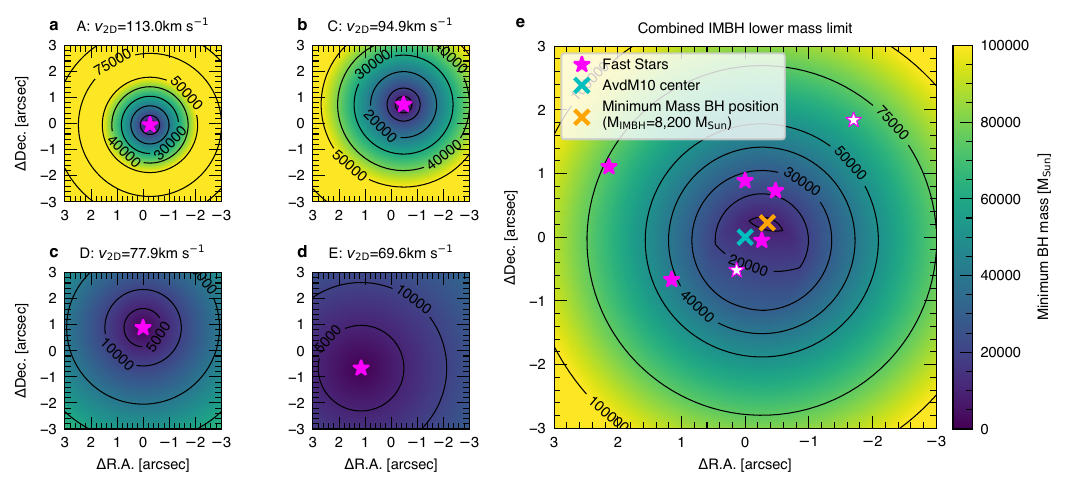}

\includegraphics[width=0.85\textwidth]{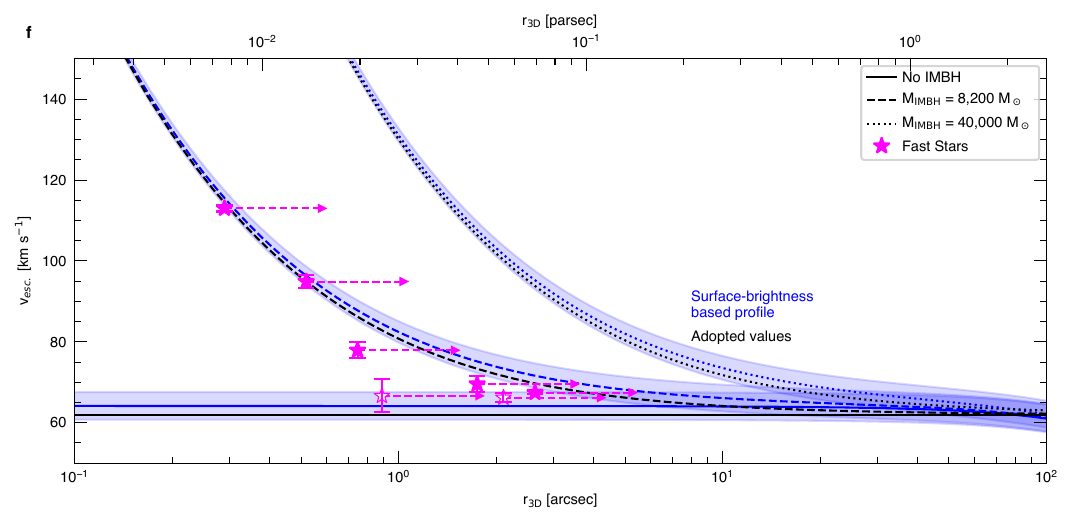}

\end{figure}

\clearpage
\edf{edf:accelerations}
\small\noindent\textbf{Extended Data Fig. \ref{edf:accelerations}: Constraints on the IMBH using the acceleration measurements. }
Taking into account the limits on the accelerations gives us additional constraints on black hole mass (\textbf{a}) and on-sky position (\textbf{b}). The contours shown correspond to the 1-, 2-, and 3-sigma levels of the distribution. This analysis using both escape velocity and acceleration measurements from the 5 robustly measured fast stars constrains the minimum IMBH mass stronger than escape velocity constraints alone. Both plots also show the distribution from an MCMC run including stars B and G.

\begin{figure}[h]
    \centering
    \includegraphics[width=1.0\textwidth]{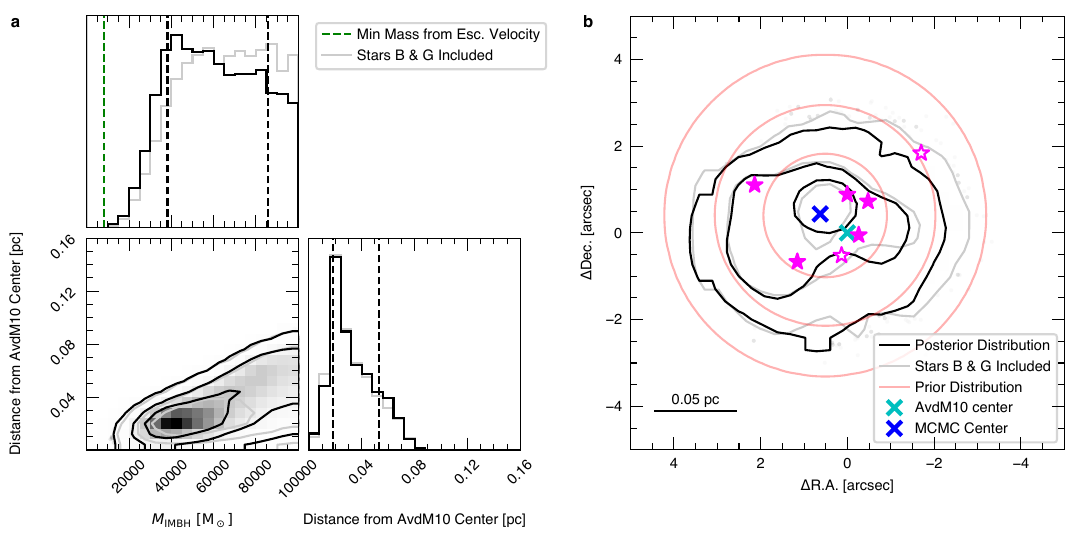}
\end{figure}

\clearpage
\edf{edf:nbody}
\small\noindent\textbf{Extended Data Fig. \ref{edf:nbody}: Comparison of the observed velocity distribution with N-Body models. }
\textbf{a}, 2D velocity distribution for the stars in the inner 10 arcseconds of $\omega$\,Cen. We show the observed data in gray, the results for an N-Body model  without an IMBH in black and the results for a model with an 47,000 M$_\odot$ IMBH in blue (based on the models of Baumgardt et al. 2019 \cite{2019MNRAS.488.5340B}). The N-Body model without an IMBH predicts no stars above the escape velocity, the 47,000 M$_\odot$ model predicts a number close to our observations. \textbf{b}, Comparison of the normalized, cumulative distribution of stellar velocities for our data and five different N-Body models.
\begin{figure}[h]%
\centering
\includegraphics[width=1.0\textwidth]{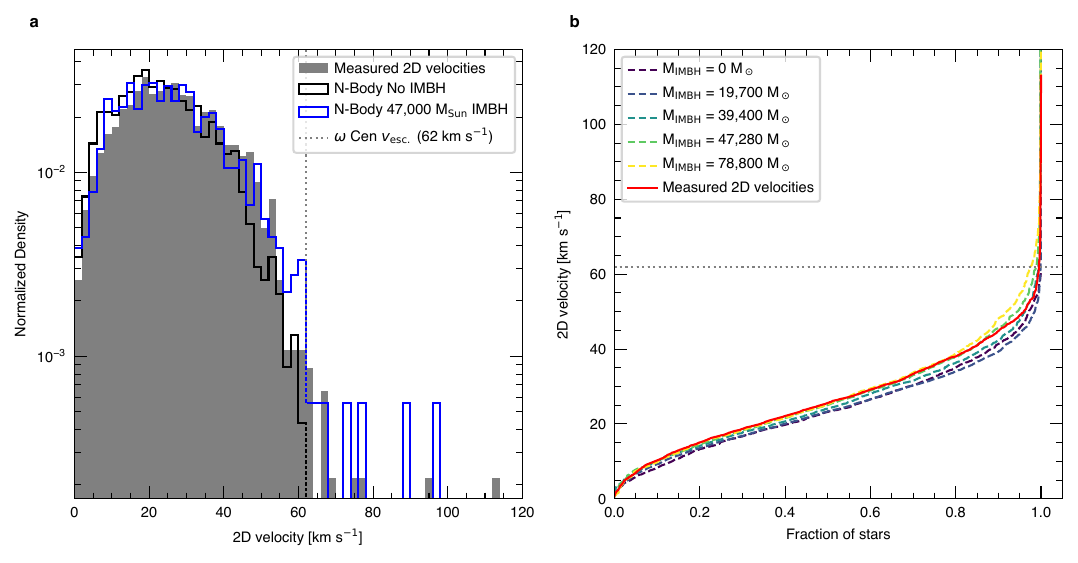}
\end{figure}
\clearpage
\edt{edf:nbody_table}
\small\noindent\textbf{Extended Data Table \ref{edf:nbody_table}: Comparison of the observed velocity distribution with N-Body models. }
In this table we report results of the comparison of the observed distribution of 2D velocities with different N-Body models  (see also Extended Data Fig. \ref{edf:nbody}). The first two columns indicate the relative and absolute mass of the IMBH in the N-Body models. The third column shows the total number of stars within 10" of the cluster center. In the fourth column, we show the results of a Kolmogorov-Smirnov test comparing the measured velocity distribution with the different N-Body models. Finally, the last three columns compare the absolute number and the fraction of fast-moving stars in the centermost 3 arcseconds.
\vspace{1cm}
\begin{center}
    \includegraphics[trim={2cm 0 2cm 10cm},width=1.0\textwidth]{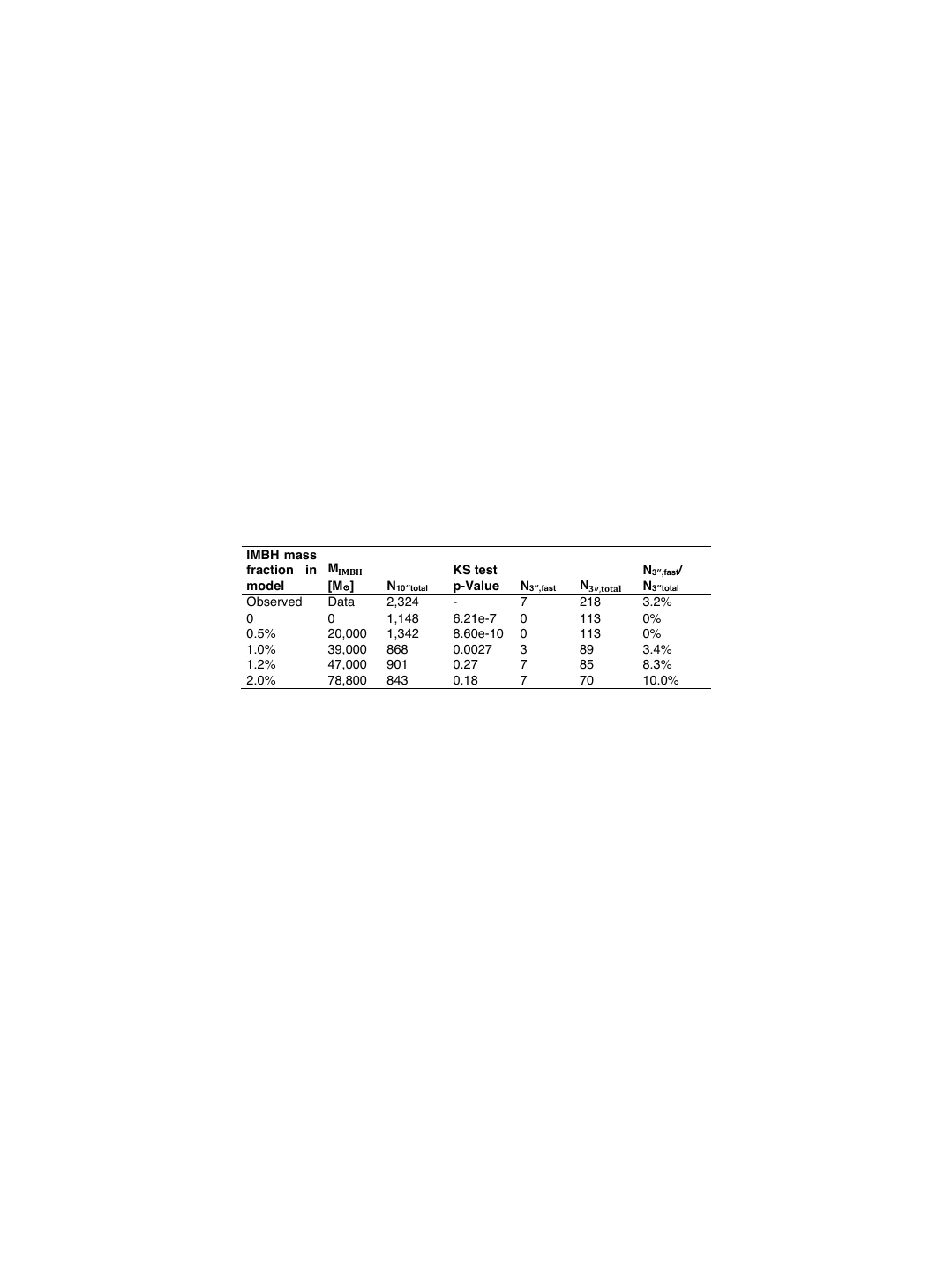}
\end{center}

\clearpage
\edf{edf:sstars}
\small\noindent\textbf{Extended Data Fig. \ref{edf:sstars}: Comparison with the Galactic Center:}
In this figure we compare the observed physical motion of our fast star sample with the stars orbiting the black hole Sgr A* in the Galactic Center \cite{2017ApJ...837...30G}. The physical scale probed by the fast-moving stars is similar to that probed by the S stars in the Milky Way center, however, the density of these tracers is lower.  Due to the approximately $\sim$100 times higher black hole mass of Sgr A*, we expect the motions to be $\sim$10 times faster and periods of the stars to be $\sim$10 times shorter, and thus show the motion for 2 years for the S stars to compare to the 20 year time span we observe the stars in $\omega$\,Cen.  
\begin{figure}[h]%
\centering
\includegraphics[width=0.45\textwidth]{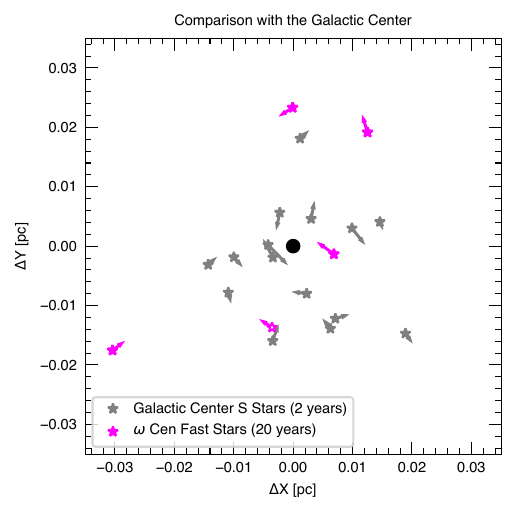}
\end{figure}

\end{document}